%
%
\magnification 1185
\baselineskip 15 pt
\tolerance=10000

\topskip .7 in
\leftskip .25 in
\rightskip .25 in
\vsize 7.2 in
\vglue .25 in

%
\def\eq#1{ \eqno({#1}) \qquad }

\def\ie{{\it i.e.\ }}
\def\ni{\noindent}
\def\Par{\par \vskip .2cm}

\font\bigggfnt=cmr10 scaled \magstep 3
 2
\font\bigfnt=cmr10 scaled \magstep  1

\font\ninerm=cmr9
\font\sevenrm=cmr7
\font\sixrm=cmr6
\font\fiverm=cmr5
\font\ninei=cmmi9
\font\sixi=cmmi6
\font\fivei=cmmi6
\font\ninesy=cmsy9
\font\sixsy=cmsy6
\font\fivesy=cmsy5
\font\tenex=cmex10
\font\nineit=cmti9
\font\ninesl=cmsl9
\font\ninett=cmtt9
\font\ninebf=cmbx9
\font\sixbf=cmbx6
\font\fivebf=cmbx5

\font\eightrm=cmr8
\font\sevenrm=cmr7
\font\sixrm=cmr6
\font\fiverm=cmr5
\font\eighti=cmmi8
\font\sixi=cmmi6
\font\fivei=cmmi6
\font\eightsy=cmsy8
\font\sixsy=cmsy6
\font\fivesy=cmsy5
\font\tenex=cmex10
\font\eightit=cmti8
\font\eightsl=cmsl8
\font\eighttt=cmtt8
\font\eightbf=cmbx8
\font\sixbf=cmbx6
\font\fivebf=cmbx5

\def\eightpoint{\def\rm{\fam0\eightrm}
  \textfont0=\eightrm \scriptfont0=\sixrm \scriptscriptfont0=\fiverm
  \textfont1=\eighti \scriptfont1=\sixi \scriptscriptfont0=\fivei
  \textfont2=\eightsy \scriptfont2=\sixsy \scriptscriptfont2=\fivesy
  \textfont3=\tenex \scriptfont3=\tenex \scriptscriptfont3=\tenex
  \textfont\itfam=\eightit  \def\it{\fam\itfam\eightit}%
  \textfont\slfam=\eightsl  \def\sl{\fam\slfam\eightsl}%
  \textfont\ttfam=\eighttt  \def\tt{\fam\ttfam\eighttt}%
  \textfont\bffam=\eightbf  \scriptfont\bffam=\sixbf
   \scriptscriptfont\bffam=\fivebf  \def\bf{\fam\bffam\eightbf}%
  \normalbaselineskip=9pt
  \setbox\strutbox=\hbox{\vrule height7pt depth2pt width0pt}%
  \let\sc=\sixrm  \normalbaselines\rm}

\def\ninepoint{\def\rm{\fam0\ninerm}
  \textfont0=\ninerm \scriptfont0=\sixrm \scriptscriptfont0=\fiverm
  \textfont1=\ninei \scriptfont1=\sixi \scriptscriptfont0=\fivei
  \textfont2=\ninesy \scriptfont2=\sixsy \scriptscriptfont2=\fivesy
  \textfont3=\tenex \scriptfont3=\tenex \scriptscriptfont3=\tenex
  \textfont\itfam=\nineit  \def\it{\fam\itfam\nineit}%
  \textfont\slfam=\ninesl  \def\sl{\fam\slfam\ninesl}%
  \textfont\ttfam=\ninett  \def\tt{\fam\ttfam\ninett}%
  \textfont\bffam=\ninebf  \scriptfont\bffam=\sixbf
   \scriptscriptfont\bffam=\fivebf  \def\bf{\fam\bffam\ninebf}%
  \normalbaselineskip=11pt
  \setbox\strutbox=\hbox{\vrule height8pt depth3pt width0pt}%
  \let\sc=\sevenrm  \normalbaselines\rm}

\def\fis{\left (\dot \phi \over \phi \right ) }

 \newbox\Ancha
 \def\gros#1{{\setbox\Ancha=\hbox{$#1$}
   \kern-.025em\copy\Ancha\kern-\wd\Ancha
   \kern.05em\copy\Ancha\kern-\wd\Ancha
   \kern-.025em\raise.0433em\box\Ancha}}


\noindent
{\bigggfnt  The initial fate of an anisotropic JBD universe } \Par 

\vskip 12 pt

\noindent
{\bigfnt H.\ N.\ N\'u\~nez-Y\'epez}{\footnote{\dag}{\rm On sabbatical leave from Departamento de F\'{\i}sica, UAM-Iztapalapa, \par e-mail: nyhn@xanum.uam.mx}} \Par

\noindent
{Instituto de F\'{\i}sica, Benem\'erita Universidad Aut\'onoma de Puebla, A\-par\-tado Pos\-tal J-48, C P 72570,  Puebla, M\'exico} \Par
\vskip 24 pt

\centerline {\bigfnt Abstract.} 
\noindent
The dynamical effects on the scale factors due to the scalar $\phi$-field at the early stages of a supposedly anisotropic Universe expansion in the scalar-tensor cosmology of Jordan-Brans and Dicke is studied. This universe shows an {\sl isotropic}  evolution  but, depending on the value of the theorie's coupling parameter $\omega$, it can  begin from a singularity if $\omega>0$ and  after expanding shrink to  another one; or, if $\omega <0$ and  $-3/2< \omega\leq -4/3$, it can evolve from a flat spatially-infinite  state to a non extended singularity; or,  if $ -4/3 < \omega < 0$, evolve from an extended  singularity  to a non singular state and, at last, proceed towards a singularity.    \Par
\vskip 12 pt

\noindent
{ Key words}:  Scalar tensor theories, Bianchi VII cosmological model. \Par 

\noindent{ PACS number(s)}: 04.20.Jb \Par 

\vfill
\hfill Por publicarse en Phys.\ Lett.\ A (1999).\Par
\eject

\noindent
Nowadays, and despite the lack  of favorable observational evidence in todays universe, there is a great deal of interest in the Jordan-Brans-Dicke scalar-tensor theory of gravity (JBD) [1, 2] ---and in other scalar tensor theories too---  because of the emergence of superstring theories which lead naturally to a dilaton theory of gravity where scalar fields are mandatory, and also due to the emergence of extended inflation models and pre-big bang cosmologies where scalar fields can provide solutions to some of the problems of inflation [3-5]. This comes about since the JBD action functional already includes a string sector where the dilaton field $\phi_D$ can be suitably related with the JBD scalar field $\phi \propto \exp(-\phi_D)$. The important role of the $\phi$-field of JBD would especially occur at the strongly relativistic stages of the Universe [5, 6]. Thence, the importance of studying the early phases of  the homogeneous and anisotropic Bianchi universes in JBD cosmology, since is possible that the early cosmological expansion of the universe be determined by an anisotropic but homogeneous vacuum stage with a non vanishing scalar field.   

In this work we report the early stages of evolution of the  Bianchi type VII$_0$ vacuum universe and show that, despite the suppposed an\-isotro\-pic behaviour of the Bianchi universes, it evolves {\sl isotropically}.   The beginnings of such universe are shown to depend on the value of the coupling parameter $\omega$ appearing in the JBD theory. If the coupling parameter is positive $\omega>0$, the universe starts from a singularity, expands reaching a maximum size  and then contracts until another singularity is reached.  If the coupling parameter is instead negative,  then the behaviour depends on whether $\omega$ is larger or smaller than the critical value $\omega_c=-4/3$. In the former case ($0>\omega>\omega_c$),  the universe begins from a spatially-infinite singularity, it quickly collapses to a finite state with non zero curvature and then, more slowly than in the previous stage,  reaches again a  singularity. In the latter case ($\omega_c>\omega>-3/2$), on the other hand, the universe starts in a spatially infinite but flat state and, from there, it collapses to a non-extended singularity. \Par

\noindent
To obtain the Bianchi-type VII  equations, let us write the line element of the spacetime, using signatu\-re $+2$  and natural units $c=G=1$, as

$$ ds^2 \, = \, -dt^2 \, + \, h_{ij}(t) \, {\gros \omega}^i \, {\gros
\omega}^j,
\eq{1} $$

\noindent
where the $h_{ij}(t)$ is the 3-metric on the surface of homogeinity, $t$ is the synchronous or cosmological time, ${\gros \omega}^i$ are the one-forms [7]:

$$ \eqalign{
{\gros \omega}^1 =& a_1 \left( (\eta  - k \nu) dy - \nu dz \right), \cr
{\gros \omega}^2 =& a_2 \left( \nu dy - (\eta  + k \nu) dz \right), \cr
{\gros \omega}^3 =& a_3 dx, \cr
{\gros \omega}^4 =& dt,
} 
\eq{2} $$

\noindent 
where $\eta \,=\, \exp (-kx) \cos(M x)$, $\nu \,=\, (-M^{-1}) \exp (-kx) \sin(M x)$, $k \,=\, h/2$ and $M \,=\, (1-k^2)^{1/2}\,$; the parameter $h$ distinguish the case analysed here ($h=0$) from the generic singular with $h\neq 0$. If we now insert the line element (1), with the forms (2), into the JBD field equations  in a vacuum [2, 8, 9], we get

$$ {\eqalign {
{d^2 \over dt^2}(\ln a_i) &+
{d\over dt} (\ln a_i) {d\over dt} (\ln a_1 a_2 a_3) +
{d\over dt} (\ln a_i)                                  \fis \cr 
&+ {\cal A}_i a_1^{-2} + {\cal E}_i \beta_2 + {\cal F}_i \beta_3 = 0,  }} \quad i={ 1,2,3}
 \eq{3} $$

\noindent 
where one of the coordinates has been chosen as the synchronous time $t$. We additionally have what we have called the constriction equation, coming from off-diagonal terms in the JBD field equations,

$$ \eqalign {
&{d\over dt}(\ln a_1) {d\over dt}(\ln a_2) +
 {d\over dt}(\ln a_1) {d\over dt}(\ln a_3) +
 {d\over dt}(\ln a_2) {d\over dt}(\ln a_3)  \cr 
& + {d\over dt}(\ln a_1 a_2 a_3) \fis - {\omega \over 2} \fis^2 +
    {\cal A}_4 a_1^{-2} + {\cal E}_4 \beta_2 + {\cal F}_4 \beta_3 = 0,
}
 \eq{4} $$

\noindent 
finally, the JBD scalar field comply with

$$ {d\over dt} \left([a_1 a_2 a_3] {d\phi\over dt} \right) = 0,
 \eq{5} $$

\noindent 
where $\beta_i \equiv (a_i / (2 a_j a_k))^2$ and the indexes $i,j,k$ are to be taken in cyclic order of 1,2,3. Equations (3), (4) and (5) are written in the standard form we introduced previously for solving the Bianchi models in JBD [10]; the specific values for the constants appearing in them are: ${\cal A}_1 = 4 M^2 - (5/2)$, ${\cal A}_2 = {\cal A}_1 - 2 $, ${\cal A}_3 = 0$, ${\cal A}_4 = {\cal A}_1 - 1$, ${\cal E}_1 = {\cal E}_3 = {\cal F}_1 = {\cal F}_2 = -2$, ${\cal E}_2 = {\cal F}_3 = 2$, ${\cal E}_4 = {\cal F}_4 = -1$, and $M^2 = 1 - (h^2 /4)$. Notice that according to these relationships, $h$ must be restricted to be $|h| \leq 2$.  The equations for the Bianchi VII$_0$ model can be obtained from these just by taking $h=0$. The Bianchi models, as consequence of certain features of their metrics, lead to additional relationships between their scale factors. In this case, we got  two additional relationships [9]

$$ {d\over dt}(\ln a_1) - {d\over dt}(\ln a_2) = 0,   
\eq{6} $$

\noindent and

$$ {h a_2 \over 2  a_1^2  a_3}  =  0. 
 \eq{7} $$

\noindent
Equation (6) always provides an additional relationship between the two scale factors, $a_1$ and $a_2$, not mattering what the value of the $h$-parameter; on the other hand, when $h\neq 0$, equation (7) leads to a singular universe in which the surfaces of homogeneity collapse to a 2-manifold, to a one-dimensional manifold or, even, to a single point. A more detailed analysis of this case is reported in [9, 11].  Equation (7) becomes just a trivial identity when $h=0$, not restricting in any way the values of the scale factors. This is precisely the case  analysed in this contribution. \par

From equation (5) we  easily obtain

$$ a_1 \, a_2 \, a_3 \, {d\phi \over dt} \,=\, \phi_0,
 \eq{8} $$

\noindent
where $\phi_0$ is an integration constant. Thus, we can define the scaled $\phi$-field, $\Phi$, as $\Phi \,\equiv\, \phi / \phi_0$, also called the intrinsic time [9, 10], and  we  get the relationship

$$ \partial_t \,=\, (a_1 \, a_2 \, a_3)^{-1} \, \partial_{\Phi}.
 \eq{9}  $$

\noindent
We  use $\Phi$ as a new time coordinate in susbstitution of  the cosmological time $t$ for solving equations (3); $\Phi$ has been found useful for analysing the Bianchi vacuum models in several situations [9, 10, 12, 13].  Let us introduce the notation $(\,)^{\prime} \,\equiv\, \partial_{\Phi}$ and, if we define the Hubble expansion rates as $H_i \,\equiv\, (\ln a_i)^{\prime}$, the field equations become

$$
H_i^{\prime} +\, {H_i \over \Phi} +\, {\cal J}_i\,a_2^4 \,+\, {\cal K}_i\,a_3^4 \,+\, {\cal N}_i\,a_2^2\, a_3^2 \,=\, 0,
\quad i={ 1,2,3}
 \eq{10} $$

\noindent
and the constriction equation becomes

$$ \eqalign {
H_1 \, H_2 \,&+\, H_1 \, H_3 \,+\, H_2 \, H_3 \,+\, {(\ln a_1 \, a_2 \,
a_3)^{\prime} \over \Phi} \,-\, {\omega \over 2 \, \Phi^2} \cr
&+\, {\cal J}_4 \, a_2^4 \,+\, {\cal K}_4 \, a_3^4 \,+\, {\cal N}_4\,
a_2^2 \, a_3^2 \,=\, 0,
}
\eq{11} $$

\noindent
the specific values for the constants appearing in equations (10) and (11) are combinations of the constants previously used: ${\cal J}_1 = {\cal J}_3 = {\cal K}_1 = {\cal K}_2 = -1/2$, ${\cal J}_2 = {\cal K}_3 = 1/2$, ${\cal J}_4 = {\cal K}_4 = -1/4$, ${\cal N}_1 = 4 M^2 - (5/2)$, ${\cal N}_2 = {\cal N}_1 - 2$, ${\cal N}_3 = 0$, ${\cal N}_4 = {\cal N}_1 - 1$.  We have written the equations as to emphasize the relationship with our previous work  on exact solutions for vacuum Bianchi models in JBD [9, 10, 12]. \Par

In the reparametrized formulation of the equations for the anisotropic homogeneous metric of Bianchi type VII$_0$, exact solutions can be obtained for the case of a Bianchi-type VII$_0$; the behaviour of such  solutions depends on the sign of the quantity 

$$\Delta \equiv - 4 ({\cal B} + 1/4), 
\eq{12}$$

\ni where, as can be readily  shown, 

$${\cal B}=\Phi^2 H_1^2 + {\Phi\over 2} H_1 - {\Phi^2 \over 2}H_1', 
\eq{13} $$ 

\noindent 
is a constant, \ie\ it is a first integral of the  system (10) and (11) expressed in terms of the Hubble expansion rates.   It can be shown that, out of the possible solutions of equations (10), the only physically plausible is the one corresponding to the case $\Delta < 0$; the cases $\Delta > 0$ or $\Delta = 0$  can be shown to led to negative or complex scale factors. For some details see [9]. The only physically admissible solution can then be explicitly written as

$$ a_1(\Phi) = \left( 4\, {\cal B} + 1 \over {c_0}^4 \right)^{1/4} \,
\left( \Phi \cosh \left[ - \sqrt{ ({\cal B} + 1/4)} \ln (f \Phi^2) \right] \right)^{-1/2},
\eq{14} $$

\noindent
where $c_0$ and $f$ are positive integration constants. The other two scale factors can be easily obtained from $a_1(\Phi)$ and  from equations (6) and (10), they are 

$$ a_2(\Phi) = c_0 \, a_1(\Phi), \qquad a_3(\Phi) =\ 2^{-1/2} c_0 a_1(\Phi). 
\eq{15} $$

\noindent
These results show that the three scale factors are proportional to each other. This means that the Bianchi VII$_0$ model, despite what we could have anticipated, shows an {\sl isotropic} evolution; it also implies that the shear $\sigma$, vorticity $\Omega$ and acceleration $\alpha$ of the reference congruence, all vanish. The vacuum Bianchi-VII$_0$ JBD universe thus basically behaves  as Friedmann-Robertson-Walker (FRW) space-time [14]. See figure 1.  The local volume on the surface of homogeneity is 

$$ V \,=\, {c_0^2 \, (a_1)^3 \over \sqrt{2}}.
\eq{16} $$

The constriction equation implies the following relationship in our case 

$${\cal B}={\omega \over 6}, 
\eq{17} $$

\noindent
 from here we notice that to have meaningful solutions the coupling parameter has to be restricted to $\omega > -3/2\equiv \omega_m$ or, ${\cal B}> -1/4$ ; $\omega_m$ is just the minimum value $\omega$ could have for the JBD theory to make sense (Ruban and Finkelstein 1975).  Notice that the specific value of $\omega$ is enough to determine the evolution of the Hubble expansion rates through equation (13), whose solution is (assuming ${\cal B}>0$)    

$$H_1(\Phi)= \sqrt {\cal B} \tanh\left(\hbox{arctanh}(H_0 \Phi_0/\sqrt{\cal B}) + 2\sqrt {\cal B}\log(\Phi_0/\Phi)\right)/\Phi; 
\eq{18}$$ 

\noindent
where $H_0\equiv H_1(\Phi_0)$. As the 3 scale factors are proportional to each other, it then follows that the three Hubble rates are always the same: $H_1=H_2=H_3$ ---another manifestation of the equivalence to a FRW isotropic universe. From (18) it can be seen that the Hubble rates, after reaching a minimum value, vanish asymptotically as $\sim-\log(\Phi)/\Phi$ when $\Phi\to \infty$.  \par

Using equations (9) and (16) we can obtain the dependence of $\Phi$ on
$t$, as follows

$$ t = { (4 {\cal B} + 1)^{3/4} \over \sqrt{2} \, {c_0}} \int \left(\Phi \cosh [ \sqrt{{\cal B} + {1/ 4}} \, \ln(f \Phi^2)]\right)^{-3/2} d \Phi,
\eq{19} $$

\noindent
though we can obtain $t$ as a function of $\Phi$, as in (19), we cannot invert it to obtain explicitly $\Phi$ as a function of $t$. Nevertheless, it is easy to realize the enormous change that occurs in $\Phi$ over a very small span of $t$ values. Behaviour of this sort has been shown to have relevance in explaining, for a Bianchi IX universe, the vanishing of the maximum Lyapunov exponent calculated in the intrinsic time $\Phi$ in spite of it being positive in the synchronous time $t$ [13]. This also shows that, asymptotically, $\Phi$ grows without bound whereas $t$ approaches a certain finite value $t_{e}$. As follows from (19), this particular value of the cosmological time can be evaluated as\par

$$ {t_e = {2\over c_0}\left((4{\cal B} +1)f^{1/3}\right)^{3/4}\int_0^{\infty} {x^{a}\over (1+x^{b})^{3/2}} dx
} \eq{20} $$\par

\noindent
where we defined $a\equiv 3(\sqrt{{\cal B} +1/4}-1/2)$, $b\equiv 4\sqrt{{\cal B} +1/4}$.  Under the conditions of the Bianchi model we are working with, expression (20) has always a $B$-dependent finite limit---which can be expressed in terms of the so-called Barnes's extended hypergeometric function [15]. See figure 2. This means, of course, that in order  to analyse the fate of the Bianchi-VII$_0$ universe for cosmological times greater than $t_e$  a different choice of coordinates is needed.  \par

 We can now state the consequences of (14) for the behaviour of the universe. An analysis of the expression (14) shows that, if $\omega >0$ (hence, only if ${\cal B} >0$), the universe  begins with a singularity since $\lim_{\Phi\to 0} a_i =0$, see equation (22), and then, as the intrinsic time unfolds, the universe expands  as $\sim \Phi^{\alpha}$, where $\alpha \equiv 3\sqrt{{\cal B} +1/4}-3/2$; the universe then reaches a maximum volume $V_{\hbox{max}}$, that can be easily calculated from (14) and (16), and then  shrinks  until it reaches again a singularity. In the other case (${\cal B}$ or $\omega <0$), as it can be ascertained from (14), (16), and (22), the fate of the universe depends of whether $\omega$ is larger than $\omega_c=-4/3$ or not. If it is  larger  than $\omega_c$ (though still negative), it is possible to show that $\lim_{\Phi\to 0} a(t)=\infty$ which coupled with the expression for the Ricci scalar (22) means that the universe starts from a non singular, spatially infinite and flat ($R=0$) state,  and from there it isotropically collapses to a singularity; that is, $\lim_{\Phi\to \infty} a(\Phi)=0$, $\lim_{\Phi\to \infty} R=\infty$. \par

 \noindent
 A universe is singular if the value of the Ricci scalar $R = g^{ab} R_{ab}$ along a certain geodesic congruence  blows up, \ie\ if $R \to \pm \infty$, whereas the associated affine parameter $s$ tends to a finite value [16]. If this happens, then we say that there exists a curvature singularity and thus that we are dealing with a singular spacetime.  For our Bianchi VII$_0$ model, we have chosen as the congruence the world lines of test observers (time-like geodesics) whose affine parameter is the synchronous time $t$. The scalar curvature can be readily  shown to be [9] 

$$ R \,=\, 2 \left( {\ddot a_1 \over a_1} + {\ddot a_2 \over a_2} +
{\ddot a_3 \over a_3} + {\ddot \phi \over \phi} + {\omega \over 2}
\left( {\dot \phi \over \phi}\right)^2 \right); 
\eq{21} $$

\noindent
using only equations (5), (10) and (11) we can rewrite $R$ for any vacuum Bianchi-VII model in terms of the coupling parameter $\omega$ and the scalar field $\phi$, as follows

$$ R \,=\, - \, \omega \, \left( {\dot \phi \over \phi} \right)^2
     \,=\, - \, \omega \, \left( {1 \over a_1 \, a_2 \, a_3 \, \Phi}
\right)^2.
\eq{22} $$

\noindent
It is important to notice that expressions (21) and (22) are valid for all the Bianchi-type VII models, not just for the specific VII$_0$ case [9]. In order than $R \to \infty$ in (21), all that is needed is that at least one the scale factor $a_1$, or just the $\phi$-field, vanish at a finite value of the synchronous time $t$; that is,  that they vanish for whatever value of  the intrinsic time $\Phi$. Notice also that we can regard the evolution of the Bianchi-type VII universe ---equations (3) and (4)--- as ``driven'' by  the curvature $R$. An analysis of the solution (14) together with (22) (basically expanding in series the expressions for $a_1(\Phi)$,) shows that, the possible singularities that may arise depend on the value of $\omega$, according to the following Table 1.

 The role of the curvature scalar in governing the universe can be qualitatively described as follows (Table 1): a) If $\omega >0$, a strong curvature prevents expansion; it is only when curvature is small that the full extent of expansion is reached but, as soon as the curvature is large in magnitude again, contraction sets in. If $0>\omega >\omega_c$, the strong curvature of the initial singular but infinite in extent state initiate the contraction until $R$ is very small, after that the universe collapses as $\Phi \to \infty$ reaching a localized singularity. If $\omega_c \geq \omega>-3/2$, the universe begins in a non-singular flat infinitely extended state and from there it collapses to a localized singularity as $\Phi \to \infty$.

 Let us conclude by saying that the supposed homogeneous and anisotropic Bianchi VII model in fact shows an isotropic expansion in the case in which $h=0$. On the other hand, we show  that the dynamics of the early stages of the expansion in this JBD model mainly depends on just one of the scale
factors (that, here, we choose as $a_1$) and that much of the universe behaviour  depends on the value of $\omega$. We have also obtained the dependence of the three scale factors $a_i$ on the intrinsic time  $\Phi$. As we have concluded that the three scale factors are proportional to each other, the expansion is necessarily isotropic and not anisotropic, as it is  usually  assumed to be in this model. In fact, we have shown that a Bianchi-VII$_0$ JBD vacuum universe is basically equivalent to a FRW-spacetime with $\sigma=0$,  $\Omega=0$ and  $\alpha=0$. 

 We think an important point of our analysis is that  shows that, even starting with supposedly anisotropic models, the inclusion of a scalar field can drastically isotropize the behaviour offering the possibility of coordinating the model with the observed isotropic properties of the actual Universe. Moreover, our analysis shows how we can change the dynamics by just changing $\omega$, a type of behaviour which gives support to the  extended inflation idea of using a dynamical coupling parameter [4, 13]. It should be clear also that from the coordinates used in this work, we cannot obtain information about what happens with the model for  synchronous times greater than $t_e$, (for example, ?`what happens as $t \to \infty$?). From this point of view, the analysis performed refers only to the early epochs of the  universe and thus it just describes its early fate.

\noindent {\bf Acknowledgements.} 

This work has been partially supported by CONACyT (grant 1343P-E9607) and by PAPIIT UNAM (grant IN122498). This work is dedicated to L.\ Bidsi, M.\ Minina, G.\ Abdul, L.\ Tuga, Q.\ Tavi, U.\ Kim, Ch.\ Cori, F.\ Cucho and G.\ Tigga for all their support and encouragement.\Par

\vfill
\eject

\noindent {\bf References}\Par

\item {[1]} Jordan, P.\ (1959)  {\sl Z.\ Phys.}, {\bf 157}, {112}. \vskip 2 pt

\item {[2]} Brans, C.\ and Dicke, R.\ H.\ (1961)  {\sl Phys.\ Rev.}, {\bf 124}, {925}. \vskip 2 pt

\item {[3]}
 Turner, M.\ S.\ Steigman, G.\ and Krauss, L.\  (1998) {\sl Phys.\ Rev.\ Lett.\ } {\bf 52}, 2090. \vskip 2pt

\item {[4]}
Steinhardt, P.\ J.\ (1993) {\sl Class.\ Quantum Grav.\  } {\bf 10} S33.\par\vskip 2pt

\item {[5]} {Gasperini, M.\ and Veneziano, G.\ (1994)} {\sl Phys.\ Rev.\ D}, {\bf 50}, {2519}. \vskip 2 pt

\item {[6]} {Clancy, D.\ Lidsey, J.\ E.\ and Tavakol, R.\ (1998)} {\sl Class.\
Quantum Grav.\ }, {\bf 15}, {257}. \vskip 2 pt

\item {[7]} {Ryan, M.\ P.\ and Shepley, L.\ C.\ (1975)} {\it Homogeous relativistic cosmologies}, {(Princeton University Press, Princeton N.\ J.\ ) pp.\ 113--267}. \vskip 2 pt

\item {[8]} {Ruban, V.\ A.\ and Finkelstein, A.\ M.\ (1975)} {\sl Gen.\ Rel.\ Grav.}, {\bf 6} {601}. \vskip 2 pt

\item {[9]} {N\'u\~nez-Y\'epez, H.\ N.\ (1995)} {\it Soluciones exactas y caos en la cosmolog\'{\i}a de Jordan, Brans y Dicke}, {Tesis doctoral, (Universidad Aut\'onoma Metropolitana, Mexico City D.\ F.\ )}.\par \vskip 2 pt

\item {[10]} {Chauvet, P.\ Cervantes-Cota, J.\ and N\'u\~nez-Y\'epez, H.\ N.\ (1992)} {\sl Class.\
Quantum Grav.\ }, {\bf 9}, 1923. \vskip 2 pt

\item {[11]} {N\'u\~nez-Y\'epez, H.\ N.\ (1999)} {\sl Astrophys.\ Space Sci.\ }, to be submitted. \vskip 2 pt

\item {[12]} {Chauvet, P.\ N\'u\~nez-Y\'epez, H.\ N.\ and Salas-Brito, A.\ L.\ (1991)} {\sl Astrophys.\ Space Sci.\ }, {\bf 178}, {165}. \vskip 2 pt

\item {[13]} {Carrretero-Gonz\'alez, R.\  N\'u\~nez-Y\'epez, H.\ N.\ and Salas-Brito, A.\ L.\ (1994)} {\sl Phys.\ Lett.\ A},  {\bf 188}, {48}. \vskip 2 pt

\item {[14]} Stephani, H.\ (1982) {\it General relativity: An introduction to the theory of the gravitational field}, (Cambridge University Press, Cambridge London). \vskip 2pt

\item {[15]} Rainville, E.\ D.\ (1980) Special Functions (Chelsea, New York)\vskip 2 pt

\item {[16]} Wald, R.\  M.\  (1984) General Relativity (University of Chicago Press, Chicago) \vskip 2pt

\vfill
\eject

\centerline{\bf Figure Captions} \Par

\ni Figure{1}\par
The behaviour of the three scale factors $a_1$, $a_2$ and $a_3$ in a Bianchi-type VII$_0$ universe against the intrinsic time is shown; these are plots of equations (14) and (15). Notice that the 3 scale factors are proportional to each other; the behaviour shown is generic in each interval of $\omega$ values. 

\noindent a)  The case ${ \omega} >0$, the especific values used for the parameters are ${\cal B}=33.333$, $c_0=2$, $f=1$. \par

\noindent b)  The case $0>\omega >\omega_c$, the values used for the parameters are ${\cal B}=-0.067$, $c_0=2$, $f=1$. \par

\noindent c)  The case $\omega_c > \omega >-3/2$, the values used for the parameters are ${\cal B}=-0.225$, $c_0=2$, $f=1$. \par
\vskip 12 pt

\ni Figure {2}\par
The maximum cosmological time $t_e$ (equation (19)) for which the analysis gives information against the value of ${\cal B} (= \omega / 6 )$.  \par \vskip 12 pt

\ni Figure {3}\par
The absolute value of the scalar curvature $|R|$ in the Bianchi-type VII$_0$ universe is shown in the range $0<|R|<\infty$ {\it versus} $\Phi$ in the range $0<\Phi<\infty$. What we really graph here is $\arctan(|R|)$ against $\arctan(\Phi)$. Small $\Phi$-values  correspond to small $t$-values but  large values of $\Phi$ correspond to $t = t_e\simeq 3.624$, as can be seen in figure 2. Giving the way the graph was compactified, you should interpret  $0$ as $0$, {\sl but\/} $\pi/2$ as $\infty$. \par

\noindent a)  The case ${ \omega} >0$, the especific values used for the parameters are ${\cal B}=33.333$, $c_0=2$, $f=1$. The universe goes from a point singularity and end in another. \par
\vskip 12 pt

\noindent b)  The case $0>\omega>\omega_c $, the values used for the parameters are ${\cal B}=-0.067$, $c_0=2$, $f=1$. \par
\vskip 12 pt

\noindent c)  The case $\omega_c >\omega >\omega_m$, the values used for the parameters are ${\cal B}=-0.225$, $c_0=2$, $f=1$. \par
\vskip 12 pt

 \vfill
 \eject

\ni {\bf Table caption}

\ni Table 1. \par
\noindent
The behaviour of the Bianchi VII$_0$ model for the different $\omega$ values is summarized in this table. $R$ is the scalar curvature. The critical value of the coupling parameter is $\omega_c= -4/3$. For $\omega\leq\omega_m\equiv -3/2$ the JBD theory loses sense. \par \vskip 12 pt

\vfill
\eject



\catcode `\!=11
\catcode `\@=11

\let\!tacr=\\ 


\newdimen\LineThicknessUnit 
\newdimen\StrutUnit            
\newskip \InterColumnSpaceUnit  
\newdimen\ColumnWidthUnit     
\newdimen\KernUnit

\let\!taLTU=\LineThicknessUnit 
\let\!taCWU=\ColumnWidthUnit   
\let\!taKU =\KernUnit          

\newtoks\NormalTLTU
\newtoks\NormalTSU
\newtoks\NormalTICSU
\newtoks\NormalTCWU
\newtoks\NormalTKU

\NormalTLTU={1in \divide \LineThicknessUnit by 300 }
\NormalTSU ={\normalbaselineskip
  \divide \StrutUnit by 11 }  
\NormalTICSU={.5em plus 1fil minus .25em}  
\NormalTCWU ={.5em}
\NormalTKU  ={.5em}

\def\NormalTableUnits{%
  \LineThicknessUnit   =\the\NormalTLTU
  \StrutUnit           =\the\NormalTSU
  \InterColumnSpaceUnit=\the\NormalTICSU
  \ColumnWidthUnit     =\the\NormalTCWU
  \KernUnit            =\the\NormalTKU}
 
\NormalTableUnits



\newcount\LineThicknessFactor    
\newcount\StrutHeightFactor      
\newcount\StrutDepthFactor       
\newcount\InterColumnSpaceFactor 
\newcount\ColumnWidthFactor      
\newcount\KernFactor
\newcount\VspaceFactor

\LineThicknessFactor    =2
\StrutHeightFactor      =8
\StrutDepthFactor       =3
\InterColumnSpaceFactor =3
\ColumnWidthFactor      =10
\KernFactor             =1
\VspaceFactor           =2


\newcount\TracingKeys 
\newcount\TracingFormats  


\def\BeginTableParBox#1{%
  \vtop\bgroup 
    \hsize=#1
    \normalbaselines 
    \let~=\!ttTie
    \let\-=\!ttDH
    \the\EveryTableParBox} 
  
\def\EndTableParBox{%
    \MakeStrut{0pt}{\StrutDepthFactor\StrutUnit}
  \egroup} 

\newtoks\EveryTableParBox
\EveryTableParBox={%
  \parindent=0pt
  \raggedright
  \rightskip=0pt plus 4em 
  \relax}


\newtoks\EveryTable
\newtoks\!taTableSpread


\newskip\LeftTabskip
\newskip\RightTabskip


\newcount\!taCountA
\newcount\!taColumnNumber
\newcount\!taRecursionLevel 

\newdimen\!taDimenA  
\newdimen\!taDimenB  
\newdimen\!taDimenC  
\newdimen\!taMinimumColumnWidth

\newtoks\!taToksA

\newtoks\!taPreamble
\newtoks\!taDataColumnTemplate
\newtoks\!taRuleColumnTemplate
\newtoks\!taOldRuleColumnTemplate
\newtoks\!taLeftGlue
\newtoks\!taRightGlue

\newskip\!taLastRegularTabskip

\newif\if!taDigit
\newif\if!taBeginFormat
\newif\if!taOnceOnlyTabskip



\def\TaBlE{%
  T\kern-.27em\lower.5ex\hbox{A}\kern-.18em B\kern-.1em
    \lower.5ex\hbox{L}\kern-.075em E}



{\catcode`\|=13 \catcode`\"=13
  \gdef\ActivateBarAndQuote{%
    \ifnum \catcode`\|=13
    \else
      \catcode`\|=13
      \def|{%
        \ifmmode
          \vert
        \else
          \char`\|
        \fi}%
    \fi
    \ifnum \catcode`\"=13
    \else
      \catcode`\"=13
      \def"{\char`\"}%
    \fi}}
 
{\catcode `\|=12 \catcode `\"=12 

}


\def\!thMessage#1{\immediate\write16{#1}\ignorespaces}
 
\let\!thx=\expandafter

\def\!thGobble#1{} 

\def\\{\let\!thSpaceToken= }\\ 

\def\!thHeight{height}
\def\!thDepth{depth}
\def\!thWidth{width}

\def\!thToksEdef#1=#2{%
  \edef\!ttemp{#2}%
  #1\!thx{\!ttemp}%
  \ignorespaces}


\def\!thStoreErrorMsg#1#2{%
  \toks0 =\!thx{\csname #2\endcsname}%
  \edef#1{\the\toks0 }}

\def\!thReadErrorMsg#1{%
  \!thx\!thx\!thx\!thGobble\!thx\string #1}

\def\!thError#1#2{%
  \begingroup
    \newlinechar=`\^^J%
    \edef\!ttemp{#2}%
    \errhelp=\!thx{\!ttemp}%
    \!thMessage{%
      ^^J\!thReadErrorMsg\!thErrorMsgA 
      ^^J\!thReadErrorMsg\!thErrorMsgB}%
    \errmessage{#1}%
  \endgroup}

\!thStoreErrorMsg\!thErrorMsgA{%
  TABLE error; see manual for explanation.}
\!thStoreErrorMsg\!thErrorMsgB{%
  Type \space H <return> \space for immediate help.}

\def\!thGetReplacement#1#2{%
   \begingroup
     \!thMessage{#1}
     \endlinechar=-1
     \global\read16 to#2%
   \endgroup}


\def\!thLoop#1\repeat{%
  \def\!thIterate{%
    #1%
    \!thx \!thIterate
    \fi}%
  \!thIterate 
  \let\!thIterate\relax}


\def\Smash{%
  \relax
  \ifmmode
    \expandafter\mathpalette
    \expandafter\!thDoMathVCS
  \else
    \expandafter\!thDoVCS
  \fi}
                      
\def\!thDoVCS#1{%
  \setbox\z@\hbox{#1}%
  \!thFinishVCS}
                      
\def\!thDoMathVCS#1#2{%
  \setbox\z@\hbox{$\m@th#1{#2}$}%
  \!thFinishVCS}
                      
\def\!thFinishVCS{%
  \vbox to\z@{\vss\box\z@\vss}}






\def\!thSetDimen{%
  \ifnum \!tgCode=1
    \ifx \!tgValue\empty
      \!taDimenA \StrutHeightFactor\StrutUnit
      \advance \!taDimenA \StrutDepthFactor\StrutUnit
      \divide \!taDimenA 2
    \else
      \!taDimenA \!tgValue\StrutUnit
    \fi
  \else
    \!taDimenA \!tgValue
  \fi
  \!taDimenA=\!thSign\!taDimenA\relax
  %
  \ifmmode
    \expandafter\mathpalette
    \expandafter\!thDoMathRaise
  \else
    \expandafter\!thDoSimpleRaise
  \fi}
                      
\def\!thDoSimpleRaise#1{%
  \setbox\z@\hbox{\raise \!taDimenA\hbox{#1}}%
  \!thFinishRaise} 
                      
\def\!thDoMathRaise#1#2{%
  \setbox\z@\hbox{\raise \!taDimenA\hbox{$\m@th#1{#2}$}}%
  \!thFinishRaise}

\def\!thFinishRaise{%
  \ht\z@\z@ 
  \dp\z@\z@
  \box\z@}


\def\!thKernBack{%
  \kern -
  \ifnum \!tgCode=1 
    \ifx \!tgValue\empty 
      \the\KernFactor
    \else
      \!tgValue    
    \fi
    \KernUnit
  \else 
    \!tgValue      
  \fi
  \ignorespaces}%

\def\Vspace{%
  \noalign
  \bgroup
  \!tgGetValue\!thVspace}

\def\!thVspace{%
  \vskip
    \ifnum \!tgCode=1 
      \ifx \!tgValue\empty 
        \the\VspaceFactor
      \else
        \!tgValue    
      \fi
      \StrutUnit
    \else 
      \!tgValue      
    \fi
  \egroup} 



  
  


\def\BeginFormat{%
  \catcode`\|=12 
  \catcode`\"=12 
  \!taPreamble={}%
  \!taColumnNumber=0
  \skip0 =\InterColumnSpaceUnit
  \multiply\skip0 \InterColumnSpaceFactor
  \divide\skip0 2
  \!taRuleColumnTemplate=\!thx{%
    \!thx\tabskip\the\skip0 }%
  \!taLastRegularTabskip=\skip0 
  \!taOnceOnlyTabskipfalse
  \!taBeginFormattrue 
  \def\!tfRowOfWidths{}
  \ReadFormatKeys}

\def\!tfSetWidth{%
  \ifx \!tfRowOfWidths \empty  
    \ifnum \!taColumnNumber>0  
      \begingroup              
         \!taCountA=1          
         \aftergroup \edef \aftergroup \!tfRowOfWidths \aftergroup {%
           \aftergroup &\aftergroup \omit
           \!thLoop
             \ifnum \!taCountA<\!taColumnNumber
             \advance\!taCountA 1
             \aftergroup \!tfAOAO
           \repeat 
           \aftergroup }%
      \endgroup
    \fi
  \fi      
  \ifx [\!ttemp 
    \!thx\!tfSetWidthText
  \else
    \!thx\!tfSetWidthValue
  \fi}

\def\!tfAOAO{%
  &\omit&\omit}

\def\!tfSetWidthText [#1]{
  \def\!tfWidthText{#1}%
  \ReadFormatKeys}

\def\!tfSetWidthValue{%
  \!taMinimumColumnWidth = 
    \ifnum \!tgCode=1 
      \ifx\!tgValue\empty 
        \ColumnWidthFactor
      \else
        \!tgValue 
      \fi
      \ColumnWidthUnit
    \else
      \!tgValue 
    \fi
  \def\!tfWidthText{}
  \ReadFormatKeys}

\def\!tfSetTabskip{%
  \ifnum \!tgCode=1
    \skip0 =\InterColumnSpaceUnit
    \multiply\skip0 
      \ifx \!tgValue\empty
        \InterColumnSpaceFactor         
      \else
       \!tgValue                        
      \fi
  \else
    \skip0 =\!tgValue                   
  \fi
  \divide\skip0 by 2
  \ifnum\!taColumnNumber=0 
    \!thToksEdef\!taRuleColumnTemplate={%
      \the\!taRuleColumnTemplate 
      \tabskip \the\skip0 }
  \else
    \!thToksEdef\!taDataColumnTemplate={%
      \the\!taDataColumnTemplate 
      \tabskip \the\skip0 }
  \fi
  \if!taOnceOnlyTabskip
  \else
    \!taLastRegularTabskip=\skip0 
  \fi                             
  \ReadFormatKeys}

\def\!tfSetVrule{%
  \!thToksEdef\!taRuleColumnTemplate={%
    \noexpand\hfil
    \noexpand\vrule
    \noexpand\!thWidth
    \ifnum \!tgCode=1
      \ifx \!tgValue\empty
        \the\LineThicknessFactor      
      \else
        \!tgValue                     
      \fi
      \!taLTU                         
    \else
      \!tgValue                       
    \fi
    ####%
    \noexpand\hfil
    \the\!taRuleColumnTemplate}       
  \!tfAdjoinPriorColumn}
 
\def\!tfSetAlternateVrule{%
  \afterassignment\!tfSetAlternateA
  \toks0 =}                           

\def\!tfSetAlternateA{%
  \!thToksEdef\!taRuleColumnTemplate={%
    \the\toks0 \the\!taRuleColumnTemplate} 
  \!tfAdjoinPriorColumn}

\def\!tfAdjoinPriorColumn{%
  \ifnum \!taColumnNumber=0
    \!taPreamble=\!taRuleColumnTemplate 
    \ifnum \TracingFormats>0             
      \!tfShowRuleTemplate
    \fi
  \else
    \ifx\!tfRowOfWidths\empty  
    \else
      \!tfUpdateRowOfWidths
    \fi
    \!thToksEdef\!taDataColumnTemplate={%
      \the \!taLeftGlue
      \the \!taDataColumnTemplate
      \the \!taRightGlue}
    \ifnum \TracingFormats>0
      \!tfShowTemplates
    \fi
    \!thToksEdef\!taPreamble={%
      \the\!taPreamble
      &
      \the\!taDataColumnTemplate
      &
      \the\!taRuleColumnTemplate}
  \fi
%
  \advance \!taColumnNumber 1
  \if!taOnceOnlyTabskip              
    \!thToksEdef\!taDataColumnTemplate={%
       ####\tabskip \the\!taLastRegularTabskip}
  \else
    \!taDataColumnTemplate{##}%
  \fi
  \!taRuleColumnTemplate{}
  \!taLeftGlue{\hfil}
  \!taRightGlue{\hfil}%
  \!taMinimumColumnWidth=0pt
  \def\!tfWidthText{}%
  \!taOnceOnlyTabskipfalse    
  \ReadFormatKeys}

\def\!tfUpdateRowOfWidths{%
  \ifx \!tfWidthText\empty
  \else 
    \!tfComputeMinColWidth
  \fi
  \edef\!tfRowOfWidths{%
    \!tfRowOfWidths
    &%
    \omit                                  
    \ifdim \!taMinimumColumnWidth>0pt
      \hskip \the\!taMinimumColumnWidth
    \fi
    &
    \omit}}                                

\def\!tfComputeMinColWidth{%
  \setbox0 =\vbox{%
    \ialign{
       \span\the\!taDataColumnTemplate\cr
       \!tfWidthText\cr}}%
  \!taMinimumColumnWidth=\wd0 }

\def\!tfShowRuleTemplate{%
  \!thMessage{}
  \!thMessage{TABLE FORMAT}
  \!thMessage{Column: Template}
  \!thMessage{%
    \space *c: ##\tabskip \the\LeftTabskip}
  \!taOldRuleColumnTemplate=\!taRuleColumnTemplate}

\def\!tfShowTemplates{%
  \!thMessage{%
    \space \space r: \the\!taOldRuleColumnTemplate}
  \!taOldRuleColumnTemplate=\!taRuleColumnTemplate
  \!thMessage{%
    \ifnum \!taColumnNumber<10
      \space
    \fi
    \the\!taColumnNumber c: \the\!taDataColumnTemplate}
  \ifdim\!taMinimumColumnWidth>0pt
    \!thMessage{%
      \space \space w: \the\!taMinimumColumnWidth}
  \fi}

\def\!tfFinishFormat{%
  \ifnum \TracingFormats>0
    \!thMessage{%
      \space \space r: \the\!taOldRuleColumnTemplate
        \tabskip \the\RightTabskip}%
    \!thMessage{%
      \space *c: ##\tabskip 0pt}
  \fi
  \ifnum \!taColumnNumber<2
    \!thError{%
      \ifnum \!taColumnNumber=0
        No
      \else
        Only 1
      \fi
      "|"}%
      {\!thReadErrorMsg\!tfTooFewBarsA
       ^^J\!thReadErrorMsg\!tfTooFewBarsB
       ^^J\!thReadErrorMsg\!tkFixIt}%
  \fi
  \!thToksEdef\!taPreamble={%
    ####\tabskip\LeftTabskip 
    &
    \the\!taPreamble \tabskip\RightTabskip
    &
    ####\tabskip 0pt \cr}
  \ifnum \TracingFormats>1
    \!thMessage{Preamble=\the\!taPreamble}
  \fi
  \ifnum \TracingFormats>2
    \!thMessage{Row Of Widths="\!tfRowOfWidths"}
  \fi
  \!taBeginFormatfalse 
  \catcode`\|=13
  \catcode`\"=13
  \!ttDoHalign}

\!thStoreErrorMsg\!tfTooFewBarsA{%
  There must be at least 2 "|"'s (and/or "\string \|"'s)}
\!thStoreErrorMsg\!tfTooFewBarsB{%
  between \string\BeginFormat\space and \string\EndFormat\space (or ".").}

\def\ReFormat[{%
  \omit
  \!taDataColumnTemplate{##}%
  \!taLeftGlue{}%
  \!taRightGlue{}%
  \catcode`\|=12  
  \catcode`\"=12  
  \ReadFormatKeys}

\def\!tfEndReFormat{%
  \ifnum \TracingFormats>0
    \!thMessage{ReF: 
       \the\!taLeftGlue
       \hbox{\the\!taDataColumnTemplate}
       \the\!taRightGlue}
  \fi
  \catcode`\|=13
  \catcode`\"=13
  \!tfReFormat}

\def\!tfReFormat#1{%
  \the \!taLeftGlue
  \vbox{%
    \ialign{%
      \span\the\!taDataColumnTemplate\cr
       #1\cr}}%
  \the \!taRightGlue}







\def\!tgGetValue#1{%
  \def\!tgReturn{#1}
  \futurelet\!ttemp\!tgCheckForParen}

\def\!tgCheckForParen{%
  \ifx\!ttemp (%
    \!thx \!tgDoParen
  \else
    \!thx \!tgCheckForSpace
  \fi}

\def\!tgDoParen(#1){%
  \def\!tgCode{2}%
  \def\!tgValue{#1}
  \!tgReturn}

\def\!tgCheckForSpace{%
  \def\!tgCode{1}%
  \def\!tgValue{}
  \ifx\!ttemp\!thSpaceToken
    \!thx \!tgReturn        
  \else
    \!thx \!tgCheckForDigit         
  \fi}

\def\!tgCheckForDigit{%
  \!taDigitfalse
  \ifx 0\!ttemp
    \!taDigittrue
  \else
    \ifx 1\!ttemp
      \!taDigittrue
    \else
      \ifx 2\!ttemp
        \!taDigittrue
      \else
        \ifx 3\!ttemp
          \!taDigittrue
        \else
          \ifx 4\!ttemp
            \!taDigittrue
          \else
            \ifx 5\!ttemp
              \!taDigittrue
            \else
              \ifx 6\!ttemp
                \!taDigittrue
              \else
                \ifx 7\!ttemp
                  \!taDigittrue
                \else
                  \ifx 8\!ttemp
                    \!taDigittrue
                  \else
                    \ifx 9\!ttemp
                      \!taDigittrue
                    \fi
                  \fi
                \fi
              \fi
            \fi
          \fi
        \fi
      \fi
    \fi
  \fi
  \if!taDigit
    \!thx \!tgGetNumber
  \else
    \!thx \!tgReturn 
  \fi}

\def\!tgGetNumber{%
  \afterassignment\!tgGetNumberA
  \!taCountA=}
\def\!tgGetNumberA{%
  \edef\!tgValue{\the\!taCountA}%
  \!tgReturn}


\def\!tgSetUpParBox{%
  \edef\!ttemp{%
    \noexpand \ReadFormatKeys
    b{\noexpand \BeginTableParBox{%
      \ifnum \!tgCode=1 
        \ifx \!tgValue\empty 
          \the\ColumnWidthFactor
        \else
          \!tgValue    
        \fi
        \!taCWU        
      \else 
        \!tgValue      
      \fi}}}%
  \!ttemp
  a{\EndTableParBox}}

\def\!tgInsertKern{%
  \edef\!ttemp{%
    \kern
    \ifnum \!tgCode=1 
      \ifx \!tgValue\empty 
        \the\KernFactor
      \else
        \!tgValue    
      \fi
      \!taKU         
    \else 
      \!tgValue      
    \fi}%
  \edef\!ttemp{%
    \noexpand\ReadFormatKeys
    \ifh@            
      b{\!ttemp}
    \fi
    \ifv@            
      a{\!ttemp}
    \fi}%
  \!ttemp}




\def\NewFormatKey#1{%
  \!thx\def\!thx\!ttempa\!thx{\string #1}%
  \!thx\def\!thx\!ttempb\!thx{\csname !tk<\!ttempa>\endcsname}%
  \ifnum \TracingKeys>0
    \!tkReportNewKey
  \fi
  \!thx\ifx \!ttempb \relax
    \!thx\!tkDefineKey
  \else 
    \!thx\!tkRejectKey
  \fi}

\def\!tkReportNewKey{%
  \!taToksA\!thx{\!ttempa}%
  \!thMessage{NEW KEY: "\the\!taToksA"}}

\def\!tkDefineKey{%
  \!thx\def\!ttempb}%

\def\!tkRejectKey{%
    \!taToksA\!thx{\!ttempa}%
    \!thError{Key letter "\the\!taToksA" already used}
      {\!thReadErrorMsg\!tkFixIt}
    \def\!tkGarbage}%

\!thStoreErrorMsg\!tkFixIt{%
  You'd better type \space 'E' \space and fix your file.}


\def\ReadFormatKeys#1{%
  \!thx\def\!thx\!ttempa\!thx{\string #1}%
  \!thx\def\!thx\!ttempb\!thx{\csname !tk<\!ttempa>\endcsname}%
  \ifnum \TracingKeys>1
    \!tkReportKey
  \fi
  \!thx\ifx \!ttempb\relax 
    \!thx\!tkReplaceKey
  \else
    \!thx\!ttempb
  \fi}

\def\!tkReportKey{%
  \!taToksA\!thx{\!ttempa}%
  \!thMessage{KEY: "\the\!taToksA"}}

\def\!tkReplaceKey{%
  \!taToksA\!thx{\!ttempa}%
  \!thError {Undefined format key "\the\!taToksA"}
    {\!thReadErrorMsg\!tkUndefined ^^J\!thReadErrorMsg\!tkBadKey}
  \!tkReplaceKeyA}

\def\!tkReplaceKeyA{%
  \!thGetReplacement{\!thReadErrorMsg\!tkReplace}\!tkReplacement
  \!thx\ReadFormatKeys\!tkReplacement}

\!thStoreErrorMsg\!tkUndefined{%
  The format key in " "'s on the next to top line is undefined.}
\!thStoreErrorMsg\!tkBadKey{%
  Type \space E \space to quit now, or
  \space<CR> \space and respond to next prompt.}
\!thStoreErrorMsg\!tkReplace{%
  Type \space<replacement key><CR> \space,
   or simply \space<CR> \space to skip offending key:}


\NewFormatKey b#1{%
  \!thx\!tkJoin\!thx{\the\!taDataColumnTemplate}{#1}%
  \ReadFormatKeys}

\def\!tkJoin#1#2{%
  \!taDataColumnTemplate{#2#1}}%

\NewFormatKey a#1{%
  \!taDataColumnTemplate\!thx{\the\!taDataColumnTemplate #1}%
  \ReadFormatKeys}

\NewFormatKey \{{%
  \!taDataColumnTemplate=\!thx{\!thx{\the\!taDataColumnTemplate}}%
  \ReadFormatKeys}

\NewFormatKey *#1#2{%
  \!taCountA=#1\relax
  \!taToksA={}%
  \!thLoop 
    \ifnum \!taCountA > 0
    \!taToksA\!thx{\the\!taToksA #2}%
    \advance\!taCountA -1
  \repeat 
  \!thx\ReadFormatKeys\the\!taToksA}


\NewFormatKey \LeftGlue#1{%
  \!taLeftGlue{#1}%
  \ReadFormatKeys}

\NewFormatKey \RightGlue#1{%
  \!taRightGlue{#1}%
  \ReadFormatKeys}

\NewFormatKey c{%
  \ReadFormatKeys 
  \LeftGlue\hfil
  \RightGlue\hfil}

\NewFormatKey l{%
  \ReadFormatKeys 
  \LeftGlue{}   
  \RightGlue\hfil}

\NewFormatKey r{%
  \ReadFormatKeys 
  \LeftGlue\hfil
  \RightGlue{}}

\NewFormatKey k{%
  \h@true
  \v@true
  \!tgGetValue{\!tgInsertKern}}

\NewFormatKey i{%
  \h@true
  \v@false
  \!tgGetValue{\!tgInsertKern}}
  
\NewFormatKey j{%
  \h@false
  \v@true
  \!tgGetValue{\!tgInsertKern}}


\NewFormatKey n{%
  \def\!tnStyle{}%
   \futurelet\!tnext\!tnTestForBracket}

\NewFormatKey N{%
  \def\!tnStyle{$}%
   \futurelet\!tnext\!tnTestForBracket}


\NewFormatKey m{%
  \ReadFormatKeys b$ a$}

\NewFormatKey M{%
  \ReadFormatKeys \{ b{$\displaystyle} a$}

\NewFormatKey \m{%
  \ReadFormatKeys l b{{}} m}

\NewFormatKey \M{%
  \ReadFormatKeys l b{{}} M}

\NewFormatKey f#1{%
  \ReadFormatKeys b{#1}}

\NewFormatKey B{%
  \ReadFormatKeys f\bf}

\NewFormatKey I{%
  \ReadFormatKeys f\it}

\NewFormatKey S{%
  \ReadFormatKeys f\sl}

\NewFormatKey R{%
  \ReadFormatKeys f\rm}

\NewFormatKey T{%
  \ReadFormatKeys f\tt}

\NewFormatKey p{%
  \!tgGetValue{\!tgSetUpParBox}}


\NewFormatKey w{%
  \!tkTestForBeginFormat w{\!tgGetValue{\!tfSetWidth}}}


\NewFormatKey s{%
  \!taOnceOnlyTabskipfalse    
  \!tkTestForBeginFormat t{\!tgGetValue{\!tfSetTabskip}}}

\NewFormatKey o{%
  \!taOnceOnlyTabskiptrue
  \!tkTestForBeginFormat o{\!tgGetValue{\!tfSetTabskip}}}


\NewFormatKey |{%
  \!tkTestForBeginFormat |{\!tgGetValue{\!tfSetVrule}}}

\NewFormatKey \|{%
  \!tkTestForBeginFormat \|{\!tfSetAlternateVrule}}


\NewFormatKey .{%
  \!tkTestForBeginFormat.{\!tfFinishFormat}} 

\NewFormatKey \EndFormat{%
  \!tkTestForBeginFormat\EndFormat{\!tfFinishFormat}} 

\NewFormatKey ]{%
  \!tkTestForReFormat ] \!tfEndReFormat}


\def\!tkTestForBeginFormat#1#2{%
  \if!taBeginFormat  
    \def\!ttemp{#2}%
    \!thx \!ttemp    
  \else
    \toks0={#1}%
    \toks2=\!thx{\string\ReFormat}%
    \!thx \!tkImproperUse
  \fi}   

\def\!tkTestForReFormat#1#2{%
  \if!taBeginFormat  
    \toks0={#1}%
    \toks2=\!thx{\string\BeginFormat}%
    \!thx \!tkImproperUse
  \else
    \def\!ttemp{#2}%
    \!thx \!ttemp    
  \fi}   

\def\!tkImproperUse{%
  \!thError{\!thReadErrorMsg\!tkBadUseA "\the\toks0 "}%
    {\!thReadErrorMsg\!tkBadUseB \the\toks2 \space command.
    ^^J\!thReadErrorMsg\!tkBadKey}%
  \!tkReplaceKeyA}
 
\!thStoreErrorMsg\!tkBadUseA{Improper use of key }  
\!thStoreErrorMsg\!tkBadUseB{%
  The key mentioned above can't be used in a }




\def\!tnTestForBracket{%
  \ifx [\!tnext
    \!thx\!tnGetArgument
  \else
    \!thx\!tnGetCode
  \fi}

\def\!tnGetCode#1 {
  \!tnConvertCode #1..!}

\def\!tnConvertCode #1.#2.#3!{%
  \begingroup
    \aftergroup\edef \aftergroup\!ttemp \aftergroup{%
      \aftergroup[%
      \!taCountA #1
      \!thLoop
        \ifnum \!taCountA>0
        \advance\!taCountA -1
        \aftergroup0
      \repeat
      \def\!ttemp{#3}%
      \ifx\!ttemp \empty
      \else
        \aftergroup.
        \!taCountA #2
        \!thLoop 
          \ifnum \!taCountA>0
          \advance\!taCountA -1
          \aftergroup0
        \repeat
      \fi 
      \aftergroup]\aftergroup}%
    \endgroup\relax
    \!thx\!tnGetArgument\!ttemp}
  
\def\!tnGetArgument[#1]{%
  \!tnMakeNumericTemplate\!tnStyle#1..!}

\def\!tnMakeNumericTemplate#1#2.#3.#4!{
  \def\!ttemp{#4}%
  \ifx\!ttemp\empty
    \!taDimenC=0pt
  \else
    \setbox0=\hbox{\m@th #1.#3#1}%
    \!taDimenC=\wd0
  \fi
  \setbox0 =\hbox{\m@th #1#2#1}%
  \!thToksEdef\!taDataColumnTemplate={%
    \noexpand\!tnSetNumericItem
    {\the\wd0 }%
    {\the\!taDimenC}%
    {#1}%
    \the\!taDataColumnTemplate}  
  \ReadFormatKeys}

\def\!tnSetNumericItem #1#2#3#4 {
  \!tnSetNumericItemA {#1}{#2}{#3}#4..!}

\def\!tnSetNumericItemA #1#2#3#4.#5.#6!{%
  \def\!ttemp{#6}%
  \hbox to #1{\hss \m@th #3#4#3}%
  \hbox to #2{%
    \ifx\!ttemp\empty
    \else
       \m@th #3.#5#3%
    \fi
    \hss}}




\def\MakeStrut#1#2{%
  \vrule width0pt height #1 depth #2}

\def\StandardTableStrut{%
  \MakeStrut{\StrutHeightFactor\StrutUnit}
    {\StrutDepthFactor\StrutUnit}}

\def\AugmentedTableStrut#1#2{%
  \dimen@=\StrutHeightFactor\StrutUnit
  \advance\dimen@ #1\StrutUnit
  \dimen@ii=\StrutDepthFactor\StrutUnit
  \advance\dimen@ii #2\StrutUnit
  \MakeStrut{\dimen@}{\dimen@ii}}

\def\Enlarge#1#2{
  \!taDimenA=#1\relax
  \!taDimenB=#2\relax
  \let\!TsSpaceFactor=\empty
  \ifmmode
    \!thx \mathpalette
    \!thx \!TsEnlargeMath
  \else
    \!thx \!TsEnlargeOther
  \fi}

\def\!TsEnlargeOther#1{%
  \ifhmode
    \setbox\z@=\hbox{#1%
      \xdef\!TsSpaceFactor{\spacefactor=\the\spacefactor}}%
  \else
    \setbox\z@=\hbox{#1}%
  \fi
  \!TsFinishEnlarge}
    
\def\!TsEnlargeMath#1#2{%
  \setbox\z@=\hbox{$\m@th#1{#2}$}%
  \!TsFinishEnlarge}

\def\!TsFinishEnlarge{%
  \dimen@=\ht\z@
  \advance \dimen@ \!taDimenA
  \ht\z@=\dimen@
  \dimen@=\dp\z@
  \advance \dimen@ \!taDimenB
  \dp\z@=\dimen@
  \box\z@ \!TsSpaceFactor{}}


\def\OpenUp#1#2{%
  \advance \StrutHeightFactor #1\relax
  \advance \StrutDepthFactor #2\relax}




\def\BeginTable{%
  \futurelet\!tnext\!ttBeginTable}

\def\!ttBeginTable{%
  \ifx [\!tnext
    \def\!tnext{\!ttBeginTableA}%
  \else 
    \def\!tnext{\!ttBeginTableA[c]}%
  \fi
  \!tnext}

\def\!ttBeginTableA[#1]{%
  \if #1u
    \ifmmode                 
      \def\!ttEndTable{
        \relax}
    \else                   
      \bgroup
      \def\!ttEndTable{%
        \egroup}%
    \fi
  \else
    \hbox\bgroup $
    \def\!ttEndTable{%
      \egroup 
      $
      \egroup}
    \if #1t%
      \vtop
    \else
      \if #1b%
        \vbox
      \else
        \vcenter 
      \fi
    \fi
    \bgroup      
  \fi
  \advance\!taRecursionLevel 1 
  \let\!ttRightGlue=\relax  
  \everycr={}
  \ifnum \!taRecursionLevel=1
    \!ttInitializeTable
  \fi}

\bgroup
  \catcode`\|=13
  \catcode`\"=13
  \catcode`\~=13
  \gdef\!ttInitializeTable{%
    \let\!ttTie=~ 
    \let\!ttDH=\- 
    \catcode`\|=\active
    \catcode`\"=\active
    \catcode`\~=\active
    \def |{\unskip\!ttRightGlue&&}
    \def\|{\unskip\!ttRightGlue&\omit\!ttAlternateVrule}%
    \def"{\unskip\!ttRightGlue&\omit&}
    \def~{\kern .5em}
    \def\\{\!ttEndOfRow}%
    \def\-{\!ttShortHrule}%
    \def\={\!ttLongHrule}%
    \def\_{\!ttFullHrule}%
    \def\Left##1{##1\hfill\null}
    \def\Center##1{\hfill ##1\hfill\null}
    \def\Right##1{\hfill##1}%
    \the\EveryTable}
\egroup

\let\!ttRightGlue=\relax  

\def\!ttDoHalign{%
  \baselineskip=0pt \lineskiplimit=0pt \lineskip=0pt %
  \tabskip=0pt
  \halign \the\!taTableSpread \bgroup
   \span\the\!taPreamble
   \ifx \!tfRowOfWidths \empty
   \else 
     \!tfRowOfWidths \cr %
   \fi}

\def\EndTable{%
  \egroup 
  \!ttEndTable}


\def\!ttEndOfRow{%
  \futurelet\!tnext\!ttTestForBlank}

\def\!ttTestForBlank{%
  \ifx \!tnext\!thSpaceToken  
    \!thx\!ttDoStandard
  \else
    \!thx\!ttTestForZero
  \fi}
  
\def\!ttTestForZero{%
  \ifx 0\!tnext
    \!thx \!ttDoZero
  \else
    \!thx \!ttTestForPlus
  \fi}

\def\!ttTestForPlus{%
  \ifx +\!tnext
    \!thx \!ttDoPlus
  \else
    \!thx \!ttDoStandard
  \fi}

\def\!ttDoZero#1{
  \cr} 

\def\!ttDoPlus#1#2#3{
  \AugmentedTableStrut{#2}{#3}%
  \cr} 

\def\!ttDoStandard{%
  \StandardTableStrut
  \cr}


 



\def\!ttAlternateVrule{%
  \!tgGetValue{\!ttAVTestForCode}}  

\def\!ttAVTestForCode{%
  \ifnum \!tgCode=2              
    \!thx\!ttInsertVrule         
  \else
    \!thx\!ttAVTestForEmpty
  \fi}

\def\!ttAVTestForEmpty{%
  \ifx \!tgValue\empty           
    \!thx\!ttAVTestForBlank
  \else
    \!thx\!ttInsertVrule         
  \fi}

\def\!ttAVTestForBlank{%
  \ifx \!ttemp\!thSpaceToken     
    \!thx\!ttInsertVrule
  \else
    \!thx\!ttAVTestForStar 
  \fi}

\def\!ttAVTestForStar{%
  \ifx *\!ttemp                  
    \!thx\!ttInsertDefaultPR     
  \else
    \!thx\!ttGetPseudoVrule       
  \fi}

\def\!ttInsertVrule{%
  \hfil 
  \vrule \!thWidth
    \ifnum \!tgCode=1
      \ifx \!tgValue\empty 
        \LineThicknessFactor
      \else
        \!tgValue
      \fi
      \LineThicknessUnit
    \else
      \!tgValue
    \fi
  \hfil
  &}

\def\!ttInsertDefaultPR*{%
  \PseudoVrule    
  &}

\def\!ttGetPseudoVrule#1{%
  \toks0={#1}%
  #1&}

\def\PseudoVrule{}

%
\def\!ttuse#1{%
  \ifnum #1>\@ne 
    \omit 
    \@multicnt=#1 
    \advance\@multicnt by \m@ne
    \advance\@multicnt by \@multicnt
    \!thLoop 
      \ifnum\@multicnt>\@ne 
      \sp@n %
    \repeat 
    \span 
  \fi}

\def\!ttUse#1[{%
  \!ttuse{#1}%
  \ReFormat[}


\def\!ttFullHrule{%
  \noalign
  \bgroup
  \!tgGetValue{\!ttFullHruleA}}

\def\!ttFullHruleA{%
  \!ttGetHalfRuleThickness 
  \hrule \!thHeight \dimen0 \!thDepth \dimen0
  \penalty0 
  \egroup} 

\def\!ttShortHrule{%
  \omit
  \!tgGetValue{\!ttShortHruleA}}

\def\!ttShortHruleA{%
  \!ttGetHalfRuleThickness 
  \leaders \hrule \!thHeight \dimen0 \!thDepth \dimen0 \hfill
  \null    
  \ignorespaces} 

\def\!ttLongHrule{%
  \omit\span\omit\span \!ttShortHrule}

\def\!ttGetHalfRuleThickness{%
  \dimen0 =
    \ifnum \!tgCode=1
      \ifx \!tgValue\empty
        \LineThicknessFactor
      \else
        \!tgValue    
      \fi
      \LineThicknessUnit
    \else
      \!tgValue      
    \fi
  \divide\dimen0 2 }



\def\WidenTableBy#1{%
  \ifdim #1=0pt
    \!taTableSpread={}%
  \else
    \!taTableSpread={spread #1}%
  \fi}

%


\def\JustLeft{%
  \omit \let\!ttRightGlue=\hfill}
\def\JustCenter{%
  \omit \hfill\null \let\!ttRightGlue=\hfill}

\let\\=\!tacr
\catcode`\!=12
\catcode`\@=12


$$ \BeginTable
\def\C{\JustCenter}
\BeginFormat
| r | r | r | r| r |
\EndFormat
\_
|\C Coupling Parameter | Initial Size | Final Size | Initial $R$ | Final $R$ | \\+22
\_

| $\omega >0$               | $0$     | $0$ | $-\infty$ | $-\infty$ | \\+20
| $\omega=0$                |  finite | $0$ | $0$       | $0$       | \\
| $0>\omega>-4/3$       | $\infty$| $0$ | $\infty$  | $\infty$  | \\
| $\omega=-4/3$         | $\infty$| $0$ | $0$       | $\infty$  | \\
| $-4/3>\omega>-3/2$| $\infty$| $0$ | $0$       | $\infty$  | \\+02
\_
\EndTable
$$
\vfill
\eject
  \end